\documentclass[12pt]{article}

\def\overc#1{\stackrel{c}{#1}}
\def\nablac{\stackrel{c}{\nabla}}
\def\r{{\cal R}}
\title{Geodesic Congruences in the Palatini $f(\r)$ Theory}
\author{Fatimah Shojai, and  Ali Shojai\\
Department of Physics, University of Tehran, Tehran, Iran.}
\date{}
\begin{document}
\maketitle
\begin{abstract}
We shall investigate the properties of a congruence of geodesics in the framework of Palatini $f(\r)$ theories. We shall evaluate the modified geodesic deviation equation and the Raychaudhuri's equation and show that $f(\r)$ Palatini theories do not necessarily lead to attractive forces. Also we shall study energy condition  for $f(\r)$ Palatini gravity via a perturbative analysis of the Raychaudhuri's equation.

\end{abstract}
\section{Introduction}
Recently modified gravity theories in which the gravitational lagrangian is an arbitrary function of the Ricci scalar ($f(\r)$ gravity\cite{soti1,new}) has received increasing attention both from the gravitational and cosmological aspects. The flat rotation curve of galaxies and the current observation of the accelerated expansion of the universe are two important motivations for such a theory\cite{soti2}.

One can categorize $f(\r)$ gravity theories in three classes. First, one may consider the metric as the only dynamical variable and assume that covariant derivatives are metric compatible, i.e. taking the connection to be the Levi-Civita connection of the metric. Such a theory is called metric $f(\r)$ theory.

Choosing the metric and the connection as independent dynamical variables, leads to the second class of such theories, usually called metric--affine $f(\r)$ theories\cite{soti1}. In this case some complexities arises as the matter action \textit{should} satisfy some consistency relations\cite{soti1,new}.

One can simplify the situation by considering a third class called Palatini $f(\r)$ theories, in which the gravitational part of the action depends both on the metric and the connection, but the matter part is independent of the affine connection. That is the metric connection is used in the matter action.

For Einstein's theory of gravity ($f(\r)=\r$) all these three classes leads to the same theory, but for a general $f(\r)$ theory they differ. The reader is refered to the literature for field equations of each theory\cite{soti1,new}, but here we shall briefly review the case of Palatini $f(\r)$ theory.

The appropriate action of the Palatini $f(\r)$ gravity is:
\begin{equation}
{\cal A}=\frac{1}{2\kappa}\int d^4x\sqrt{-g}f(\r[g,\Gamma])+{\cal A}_m
\label{action}
\end{equation}
where $\kappa=8\pi G/c^4$, and ${\cal A}_m$ is the matter action and has no dependence on the connection. Varying action (\ref{action}) with respect to the metric ($g_{\mu\nu}$) and 
the connection ($\Gamma^\lambda_{\mu\nu}$) yields to the following field equations:
\begin{equation}
f'(\r)\r_{\mu\nu}-\frac{1}{2}f(\r)g_{\mu\nu}=\kappa T_{\mu\nu}
\label{mee}
\end{equation}
and 
\begin{equation}
\nabla_\gamma(f'(\r)\sqrt{-g}g^{\mu\nu})=0
\label{sevom}
\end{equation}
where we have assumed that the connection is symmetric and $\nabla_\gamma$ indicates covariant derivative with respect to affine connection.
By the latter equation one has:
\begin{equation}
\Gamma^\lambda_{\mu\nu}={\lambda\brace\mu\nu}+\Delta^\lambda_{\mu\nu}
\label{af}
\end{equation} 
where:
\begin{equation}
\Delta^\alpha_{\beta\gamma}=\frac{1}{2}(\delta^\alpha_\beta\partial_\gamma\ln f'+\delta^\alpha_\gamma\partial_\beta\ln f'-g_{\beta\gamma}g^{\alpha\delta}\partial_\delta\ln f')
\label{aaf}
\end{equation}
is the difference between the affine connection and the Christoffel symbols (${\lambda\brace\mu\nu}$) and a prime denotes differentiation with respect to $\r$. 

In this paper the behaviour of a geodesics congruence in the Palatini $f(\r)$ theories is investigated. Since in these theories one deals with two different connections (Christoffel symbols and the affine connection),  what is meant by (for example) \textit{geodesic} and \textit{geodesic deviation} should be clarified. It is a well-known fact that in the Palatini $f(\r)$ theories, the matter energy-momentum tensor is divergence free with respect to the covariant  derivative defined with Levi-Civita connection of the metric. This is because of the fact that the matter does not couple to the connection. This implies that test particles shall move on the \textit{metric geodesics}\cite{tomi},  calculated using the Levi-Civita connection. This result can be obtained also from the variational principle.
 On the other hand the distance between neighbouring geodesics,  involves the Riemann tensor calculated from the affine connection.

More precisely, in the Palatini $f(\r)$ gravity the affine connection is not coupled to the matter, ($\frac{\delta {\cal A}_m}{\delta\Gamma^\alpha_{\beta\gamma}}=0$), and this theory is dynamically equivalent to a scalar-tensor theory with the Brans-Dicke parameter,  $\omega_0=-3/2$ \cite{soti3}. However these two theories are not completely equivalent since Palatini theory is genuinely a metric-affine theory and it  is different from a metric theory in which the connection of the space-time is the Levi-Civita connection. In order to stress this point, we recall that in a metric-affine theory the role of the affine-connection is not only in the equations of motion for metric and connection, but also it defines parallel transport and covariant derivatives. Therefore different connections leads to the different space-time structures.  This point that the test particles move on metric geodesic, (and not on the affine curve) only means that the particle's trajectory is not a curve along which the tangent vector, particle's velocity, is propagated parallel to itself. The distance between geodesics and the description of a congruence of them are of course given by the affine connection.

To state this important point in another way, let us to stress that the theory is not only the field equations. It is the field equations derived from the action defined on some space--time with predefined properties. For the Palatini $f(\r)$ theory one assumes the space--time has an independent affine connection and thus any parallel transport should be evaluated using it. The fact that the field equations are equivalent to the Brans--Dicke theory does not means that these theories are identical. Because the Brans--Dicke theory is defined on a space--time that parallel transport is done by the Christoffel symbols of the metric.

According to the above, one expects to have changes in the geometrical concepts like geodesic deviation and Raychaudhuri's equation representing how a flux of geodesics expands. Here we shall look for the way these concepts differ from the general relativity.

\section{Geodesic Deviation in Palatini $f(\r)$ Theory}
In the Einstein's theory of gravitation the physical meaning of the Riemann tensor is illustrated by examining the behaviour of neighbouring geodesics, the geodesic deviation concept. The Riemann tensor as a geometrical object is related to the tidal forces as a physical concept. It is a good idea and seems necessary to investigate the same problem in the framework of Palatini $f(\r)$ theory.

For this propose, as for the standard general relativity, we consider a 2-surface $S$ covered by a congruence of time--like geodesics. The parametric equation of the surface is given by $x^{\alpha}(\tau,\nu)$ in which $\tau$ is an affine parameter along the specified geodesic and $\nu$ labels distinct geodesics. At any point of $S$ there exist two vector fields: $u^{\mu}=\frac{\partial{x^\mu}}{\partial{\tau}}$, $\xi^{\mu}=\frac{\partial{x^\mu}}{\partial{\nu}}$. The first one is tangent to the geodesics at that point and the second connects two nearby curves in the congruence. Therefore one has the Lie derivative relations: ${\cal L}_{u}{\xi^\mu}={\cal L}_{\xi}{u^\mu}=0$.
Since in our formulation the affine connection is symmetric as well as the Christoffel symboles, the above relations can be written either as:
\begin{equation}
u^\beta\nabla_\beta \xi^\alpha=\xi^\beta\nabla_\beta u^\alpha
\label{aval}
\end{equation}
or as:
\begin{equation}
u^\beta\nablac_\beta \xi^\alpha=\xi^\beta\nablac_\beta u^\alpha
\end{equation}
where $\nablac$ is the covariant derivative with the Christoffel symbols as the connection.

In the Palatini $f(\r)$ theory, using the N\"oether theorem one has the conservation of energy--momentum tensor in terms of the metric connection\cite{tomi}. That is to say, the geodesic equation is 
\begin{equation}
u^\mu\nablac_\mu u^\nu=0
\label{charom}
\end{equation}
These equations can be combined to prove that $\xi^\alpha u_\alpha$ and $u^\alpha u_\alpha$ are constant along any geodesics. This means that in the Palatini $f(\r)$ theory the character of a particle is invariant on the particle's trajectory. That is to say, a time-like trajectory remains time-like during the particle's motion. The same is true for space-like and light-like trajectories.  Also by an appropriate parametrization of the geodesics, $\xi^\alpha$ is everywhere orthogonal to $u^\alpha$ and so $\xi^\alpha$ can be interpreted as the deviation vector. 

In order to obtain the geodesic deviation, one has to evaluate the relative acceleration of neighbouring geodesics. This can be achieved by parallel transporting $D\xi^\alpha/D\tau=u^\mu\nabla_\mu \xi^\alpha$ which is the covariant derivative of $\xi^\alpha$ along a curve of congruence. Using relation (\ref{aval}) we obtain:
\[ 
\frac{D^2\xi^\alpha}{D\tau^2}=u^\beta\nabla_\beta(\xi^\gamma\nabla_\gamma u^\alpha)=
\]
\begin{equation}
\xi^\beta\nabla_\beta(u^\gamma\nabla_\gamma u^\alpha)-\r^\alpha_{\mu\beta\gamma} u^\mu \xi^\beta u^\gamma
\end{equation}
Although the first term vanishes in the Einstein's theory by virtue of the geodesic equation, it is not zero in the Palatini $f(\r)$ theory. Substituting $\nabla$ in terms of $\nablac$, using the relation (\ref{af}) the first term can be expressed as:
\[
\xi^\beta\nabla_\beta(u^\gamma\nabla_\gamma u^\alpha)=\xi^\beta\nabla_\beta(u^\gamma\Delta^\alpha_{\gamma\delta} u^\delta)=
\]
\begin{equation}
\frac{1}{2}\xi^\gamma\nabla_\gamma (2 u^\alpha u^\delta \partial_\delta\ln f'-g^{\alpha\epsilon}\partial_\epsilon\ln f') 
\end{equation}
And finally after calculating the above derivative one gets:
\[
\frac{D^2\xi^\alpha}{D\tau^2}=\r^\alpha_{\beta\gamma\delta} u^\beta u^\gamma \xi^\delta +
\left ( u^\alpha\frac{D\xi^\beta}{D\tau}+u^\beta\frac{D\xi^\alpha}{D\tau} \right ) \nabla_\beta\ln f' +
\]
\begin{equation}
\xi^\gamma(u^\alpha u^\beta-\frac{1}{2}g^{\alpha\beta}) \nabla_\gamma\nabla_\beta\ln f' -\frac{1}{2}\xi^\gamma g^{\alpha\beta}\nabla_\gamma\ln f'\nabla_\beta\ln f'
\end{equation}
The first term is the standard one, while the second term introduces a new concept in the geodesic deviation. It is proportional to the relative velocity. The last two terms are proportional to the relative distance and thus have the same effect as the first term. Although the first term in the standard theory provides a negative relative acceleration (so that gravity is attractive), here this is not necessarily true and thus one may have \textit{antigravity} or repulsive gravity. 
There are cases in the standard gravity for which the gravity is repulsive. For example, for Israel shells, there is a repulsive gravitational effect which depends on the acceleration of the shell observers\cite{isper}. Also it is possible to have repulsive gravitational fields in the presence of domain walls\cite{mon}. Circumstances under which the gravity might be repulsive are studied in \cite{gas}.
But the root of repulsive gravity here, is different. It is a result of both having $f(\r)$ instead of $\r$ as the Lagrangian density and also having a connection different from the Christoffel's symbols.
We shall see in  the next section explicitly that even if the conventional energy conditions are satisfied, it is possible to have repulsive gravity.
\section{Raychaudhuri's Equation in Palatini $f(\r)$  Theory}
In order to investigate the relation between the nearby geodesics more precisely one can use the Raychaudhuri's equation. In the Einstein's theory of gravity, the role of the Raychaudhuri's equation is to guarantee that gravity acts always as an attractive force, provided the strong energy condition is satisfied\cite{wald}.
To see what differences arise for our case, in this section firstly we discuss the kinematics of a congruence of geodesic and then we shall derive the Raychaudhuri's equation for the Palatini $f(\r)$ gravity. Much of the techniques are parallel to the case of Einstein's gravity.
\subsection{Kinematics of a congruence of timelike geodesics}
Consider the same geometrical setup of the previous section. Let us introduce the tensor field, $\nabla_\beta u^\alpha$, which can be expressed as:
\begin{equation}
\nabla_\beta u^\alpha=g^{\alpha\gamma}B_{\gamma\beta}+\Delta^\alpha_{\beta\delta}g^{\delta\gamma}u_\gamma
\end{equation}
where $B_{\gamma\beta}\equiv\nablac_{\beta}u_\gamma$\cite{poisson} is a purely transverse tensor to the congruence of the geodesics. To determine the evolution of the deviation vector we need to calculate:
\begin{equation}
\frac{D\xi^\alpha}{D\tau}=u^\beta\nabla_\beta\xi^\alpha=\xi^\beta\nabla_\beta u^\alpha\equiv \tilde{B}^\alpha_\beta\xi^\beta
\end{equation}
where
\begin{equation}
\tilde{B}^\alpha_\beta=B^\alpha_\beta+\frac{1}{2}u^\alpha\partial_\beta\ln f'+\frac{1}{2}\frac{D\ln f'}{D\tau}\delta^\alpha_\beta
\end{equation}
Therefore $\tilde{B}^\alpha_\beta$ determines the evolution of the deviation vector. It has to be noted that it is not purely transverse unlike $B^\alpha_\beta$. To understand the geometric interpretation of this tensor, we can decompose it into its spherical tensor parts, its trace, a traceless symmetric tensor, and an antisymmetric tensor:
\begin{equation}
\tilde{B}_{\alpha\beta}=\frac{1}{3}\tilde{\theta}g_{\alpha\beta} + \tilde{\sigma}_{\alpha\beta} + \tilde{\omega}_{\alpha\beta}
\end{equation}
where
\begin{equation}
\tilde{\theta}=\theta+\frac{15}{2}\frac{D\ln f'}{D\tau}
\label{dovom1}
\end{equation}
\begin{equation}
\tilde{\sigma}^\alpha_\beta=\sigma^\alpha_\beta-2\frac{D\ln f'}{D\tau}\delta^\alpha_\beta+\frac{1}{4} (u^\alpha \nabla_\beta+u_\beta\nabla^\alpha)\ln f'
\label{dovom2}
\end{equation}
\begin{equation}
\tilde{\omega}^\alpha_\beta=\omega^\alpha_\beta+\frac{1}{4} (u^\alpha \nabla_\beta-u_\beta\nabla^\alpha)\ln f'
\label{dovom3}
\end{equation}
$\tilde{\theta}$, $\tilde{\sigma}^\alpha_\beta$ and $\tilde{\omega}^\alpha_\beta$ are the expansion scalar, the shear tensor and the rotation tensor respectively. The quantities without $\tilde{}$ are constructed from $B^\alpha_\beta$ in the same manner and so they are purely transverse.

Considering the congruence of the geodesics as a deformable medium, one can find the geometrical meanings of these quantities.  Consider a small displacement from one spacelike hypersurface to another one, which leads to:
\begin{equation}
\Delta\xi^\alpha=\tilde{B}^\alpha_\beta\xi^\beta(t_0)\Delta t
\end{equation}
Three cases are distinguishable:
\begin{itemize}
\item{a)} If $\tilde{\omega}^\alpha_\beta=\tilde{\sigma}^\alpha_\beta=0$:
\begin{equation}
\Delta\xi^\alpha=\frac{1}{3}\tilde{\theta}\xi^\alpha(t_0)\Delta t
\end{equation}
This shows that $\tilde{\theta}$ represents the expansion of the congruence of the geodesics. Because if one makes the two nearby geodesics synchronized at $t_0$ (that is $\xi^0(t_0)=0$), they remain synchronized at $\Delta t$ seconds later and the spatial distance expands at a rate proportional to $\tilde{\theta}$ and $\vec{\xi}(t_0)$. Also the relation (\ref{dovom1}) shows that the expansion parameter depends on the choice of arbitrary function $f(\r)$.  
\item{b)} If $\tilde{\sigma}^\alpha_\beta=0$ and $\tilde{\theta}=0$:
\begin{equation}
\Delta\xi^\alpha=\tilde{\omega}^\alpha_\beta\xi^\beta(t_0)\Delta t
\end{equation}
This leads to some asynchronization:
\begin{equation}
\Delta\xi^0=\tilde{\omega}^0_i\xi^i\Delta t
\end{equation}
where $\tilde{\omega}^0_i=\frac{1}{4}(u^0\nabla_i-u_i\nabla^0)\ln f'$, and rotation of the congruence:
\begin{equation}
\Delta\vec{\xi}=\vec{\Omega}\times\vec{\xi}\Delta t
\end{equation}
where $\epsilon_{ijk}\Omega^k=\tilde{\omega}_{ij}$. We see that although $\tilde{\omega}^i_j$ is concerned with the overall rotation of the congruence, like it's role in general relativity, we have also some asynchronization produced by $\tilde{\omega}^0_i$ elements. 
\item{c)} If $\tilde{\omega}^\alpha_\beta=0$ and $\tilde{\theta}=0$:
\begin{equation}
\Delta\xi^\alpha=\tilde{\sigma}^\alpha_\beta\xi^\beta(t_0)\Delta t
\end{equation}
Again we have some asynchronization:
\begin{equation}
\Delta\xi^0=\tilde{\sigma}^0_i\xi^i\Delta t
\end{equation}
where $\tilde{\sigma}^0_i=\frac{1}{4}(u^0\nabla_i+u_i\nabla^0)\ln f'$, and shearing of the congruence:
\begin{equation}
\Delta\xi^j=\tilde{\sigma}^j_i\xi^i\Delta t
\end{equation}
As a result a 3 sphere would deform to an ellipsoid with its axis as the principal axis of the spatial part of $\tilde{\sigma}$.
\end{itemize}
\subsection{Raychaudhuri's equation}
Now we want to derive an evolution equation for the expansion scalar. We begin by evaluating the time derivative of $\frac{D\tilde{B}^\alpha_\beta}{D\tau}$ and then substituting $\nabla$ in terms of $\nablac$ using equation (\ref{af}) and (\ref{aaf}),we arrive at:
\[
u^\gamma\nabla_\gamma\tilde{B}_{\alpha\beta}=u^\gamma\nablac_\gamma B_{\alpha\beta} - u^\gamma \left (\Delta^\delta_{\gamma\alpha}B_{\delta\beta}+\Delta^\delta_{\gamma\beta}B_{\alpha\delta}\right ) +
\]
\[
\frac{1}{2}u^\gamma\nablac_\gamma\left ( u_\alpha\nablac_\beta\ln f'\right ) -\frac{1}{2}u^\gamma \left ( \Delta^\delta_{\gamma\alpha}u_\gamma\nablac_\beta\ln f' + \Delta^\delta_{\gamma\beta}u_\alpha\nablac_\delta\ln f' \right )+
\]
\begin{equation}
\frac{1}{2}u^\gamma\nablac_\gamma\left ( u^\delta\nabla_\delta\ln f' \right )g_{\alpha\beta} +\frac{1}{2}u^\gamma\left ( u^\delta\nablac_\delta\ln f'\right )\nabla_\gamma g_{\alpha\beta}
\end{equation}
The equation for the expansion scalar is obtained by taking the trace of the above equation. And after doing some calculation one gets:
\[
\frac{D\tilde{\theta}}{D\tau}=-\frac{1}{3}\theta^2-\sigma^{\alpha\beta}\sigma_{\alpha\beta} + \omega^{\alpha\beta}\omega_{\alpha\beta}-\overc{\r}_{\alpha\beta}u^\alpha u^\beta -\frac{3}{2}\left ( \frac{D\ln f'}{D\tau}\right )^2-
\]
\begin{equation}
(\tilde\theta+\theta)\frac{D\ln f'}{D\tau} +\frac{5}{2}\frac{D^2\ln f'}{D\tau^2}
\label{rraayy}
\end{equation}
where $\overc{\r}_{\alpha\beta}$ is the Ricci tensor constructed from Levi-Civita connection. Now using the equations (\ref{dovom1}), (\ref{dovom2}) and (\ref{dovom3}) we can express the above relations with respect to the tensors describing the congruence behaviour in Palatini $f(\r)$ theory, $\tilde{\theta}$, $\tilde{\sigma}^\alpha_\beta$, $\tilde{\omega}^\alpha_\beta$ and also expressing  
$\overc{\r}_{\alpha\beta}$ with respect $\r_{\alpha\beta}$. The result is the modified Raychaudhuri's equation:
\[
\frac{D\tilde{\theta}}{D\tau}=-\frac{1}{3}\tilde{\theta}^2-\tilde\sigma^{\alpha\beta}\tilde\sigma_{\alpha\beta} + \tilde\omega^{\alpha\beta}\tilde\omega_{\alpha\beta}-\r_{\alpha\beta}u^\alpha u^\beta -27\left ( \frac{D\ln f'}{D\tau}\right )^2+
\]
\begin{equation}
3\tilde\theta\frac{D\ln f'}{D\tau} +\frac{3}{2}\frac{D^2\ln f'}{D\tau^2} -\frac{1}{2}\nabla_\alpha\nabla^\alpha\ln f' +\frac{1}{2} \nabla^\alpha\ln f' \nabla_\alpha\ln f'
\label{ray}
\end{equation}
Since $\tilde{B}^\alpha_\beta$ is not purely transverse, having a hypersurface orthogonal congruence does not mean that the rotation tensor vanishes, $\tilde\omega^{\alpha\beta}\neq 0$. Therefore for such a congruence the term $\tilde{\omega}^{\alpha\beta}\tilde{\omega}_{\alpha\beta}$ is not zero. Also the term $\tilde{\sigma}^{\alpha\beta}\tilde{\sigma}_{\alpha\beta}$ is not necessarily non-negative since $\tilde{\sigma}^\alpha_\beta$ is not transverse. 

In addition, let the conventional strong energy condition,$({\cal T}_{\alpha\beta}-\frac{1}{2}{\cal T}g_{\alpha\beta})u^\alpha u^\beta\geq 0$, holds, we can see that  $\r_{\alpha\beta}u^\alpha u^\beta$ is not necessarily non-negative. Consider the modified Einstein's equation (\ref{mee}), its trace is:
\begin{equation}
f'\r-2f=\kappa{\cal T}
\label{tr}
\end{equation}
Combining equations (\ref{mee}) and (\ref{tr}), we get:
\begin{equation}
{\cal T}_{\alpha\beta}-\frac{1}{2}{\cal T}_{\alpha\beta}=f'\r_{\alpha\beta}-\frac{1}{2}g_{\alpha\beta}(f'\r-f)
\end{equation}
Therefore the conventional strong energy condition results in:
\begin{equation}
\r_{\alpha\beta}u^\alpha u^\beta\geq \frac{1}{2}\left ( \frac{f}{f'}-\r\right )
\end{equation}
Note that for $f=\r$ these recovers the standard relation.

It must be noted that the extra terms in the modified Raychaudhuri's equation are also not negative necessarily. So that the expansion scalar does not decrease during the evolution of the congruence. This clearly shows that in Palatini $f(\r)$ theories the gravitational force is not necessarily attractive.
\section{Conclusion and Remarks}
Although Palatini $f(\r)$ theory is one way to deviate from general relativity giving expansive behaviour in cosmology, it has some unappealing characteristics. For Pallatini $f(\r)$ gravity the Cauchy problem is not well formulated yet\cite{new,khodabereferreemargbede} and for generic choices of  $f(\r)$ there is no satisfactory physical solutions describing stars and compact objects\cite{a}. These are the present problems facing Pallatini $f(\r)$ gravity.

In spite of these, from the theoretical point of view it is instructive to study the behaviour of the geodesics in such a theory as it can present a different viewpoint for the phenomenology used in cosmology and also clarifies the fundamental theoretical character of such a theory. 
 
In this paper we have presented an analysis of the equation of the geodesic deviation and the properties of a congruence of geodesics (using the Raychaudhuri's equation), for the Palatini theory of gravity in which the gravitational lagrangian is an arbitrary function of the Ricci scalar. In this theory the existence of two different connection fields has new consequences. The geodesics are determined by the Christoffel symbols (this choice is motivated by energy-momentum conservation) but the equation that governs the evolution of the deviation vector involves the affine connection (motivated by the fact that the covariant derivative or parallel propagation along any arbitrary curve is defined by the affine connection). We have formulated the kinematics of a congruence of geodesics in terms of a tensor which is called $\tilde{B}_{\alpha\beta}$.

For Einstein's gravity theory, using the strong energy condition, one sees that the geodesic congruence is converging. But for the Palatini $f(\r)$ gravity, the Raychaudhuri's equation is so highly dependent on the Ricci scalar that one can have both divergent and convergent congruences.

Although we have dealt with the Palatini $f(\r)$ theory here, similar calculations can be carried out for a general metric-affine theory. For such a theory test particles do not follow the geodesics of either metric or the connection necessarily. Three classes of curves, free fall trajectory, metric geodesics and affine geodesics are distinguishable in metric-affine theory. Therefore one expects to have similar results for the metric--affine theories.

Furthermore in a general metric-affine theory, there are two tensors describing the matter, energy-momentum tensor (${\cal T}_{\mu\nu}\equiv -\frac{2}{\sqrt{-g}}\frac{\delta{\cal A}_m}{\delta g^{\mu\nu}}$) and the hypermomentum tensor ($\Delta_\lambda^{\mu\nu}\equiv -\frac{2}{\sqrt{-g}}\frac{\delta{\cal A}_m}{\delta \Gamma^\lambda_{\mu\nu}}$). This means that all the information about matter is not in the energy-momentum tensor. Consequently the energy conditions may have a different form for this ${\cal T}_{\alpha\beta}$.

Turning back to the Palatini case, and looking at Raychaudhuri's equation (\ref{ray}), it is obvious that the appropriate energy conditions for having convergent congruences is different from that of Einstein's gravity. To study these conditions more precisely let us to use Raychaudhuri's equation 
 (\ref{rraayy}) written in terms of the purely transverse parts of $B_{\alpha\beta}$. If we substitute $\theta$ and  $\overc{\r}_{\alpha\beta}$ in terms of  $\tilde{\theta}$ and  $\r_{\alpha\beta}$ in the right hand side of equation (\ref{rraayy}), we arrive at:
\[
\frac{D\tilde{\theta}}{D\tau}=-\frac{1}{3}\tilde{\theta}^2-\sigma^{\alpha\beta}\sigma_{\alpha\beta} + \omega^{\alpha\beta}\omega_{\alpha\beta}-\r_{\alpha\beta}u^\alpha u^\beta -\frac{41}{4}\left ( \frac{D\ln f'}{D\tau}\right )^2+
\]
\begin{equation}
3\tilde\theta\frac{D\ln f'}{D\tau} +\frac{3}{2}\frac{D^2\ln f'}{D\tau^2} -\frac{1}{2}\nabla_\alpha\nabla^\alpha\ln f' 
\label{khob}
\end{equation}

Now consider a hypersurface orthogonal congruence. According to the Frobenius' theorem \cite{wald}, the purely transverse part of the rotation tensor, $\omega_{\alpha\beta}$ is zero. Also since the transverse part of the shear tensor, $\sigma_{\alpha\beta}$, is purely spatial, the second term in the equation (\ref{khob}) is non-negative. Therefore one can be sure about the convergence of a congruence of timelike geodesic by requiring:
\begin{equation}
\r_{\alpha\beta}u^\alpha u^\beta +\frac{41}{4}\left ( \frac{D\ln f'}{D\tau}\right )^2-
3\tilde\theta\frac{D\ln f'}{D\tau} -\frac{3}{2}\frac{D^2\ln f'}{D\tau^2}+\frac{1}{2}\nabla_\alpha\nabla^\alpha\ln f' \geq 0
\label{ohoy}
\end{equation}

Let us proceed with the case that the theory differs slightly from the Einstein's theory. We can cope with the method for metric $f(\r)$ theory\cite{kung}. Writing $f(\r)= \r+\beta{\cal S(\r)}$, where $\beta$ is a small parameter, the field equation (\ref{mee}) leads to:
\begin{equation}
(1+\beta {\cal S}'){\cal R}_{\mu\nu}-\frac{1}{2}({\cal R}+\beta {\cal S})g_{\mu\nu}=\kappa {\cal T}_{\mu\nu}
\label{six}
\end{equation}
The trace of this equation up to first order in $\beta$ is:
\begin{equation}
{\cal R}+2\beta {\cal S}+\kappa {\cal T}(1+\beta {\cal S}')=0
\label{seven}
\end{equation}
Combining equations (\ref{six}) and (\ref{seven}) we also get:
\begin{equation}
{\cal R}_{\mu\nu}=\kappa \left ({\cal T}_{\mu\nu}-\frac{1}{2}{\cal T}g_{\mu\nu}\right ) -\beta \left (\kappa {\cal T}_{\mu\nu}{\cal S}'+\frac{1}{2}{\cal S}g_{\mu\nu}\right )
\label{eight}
\end{equation}
Keeping only first order terms, we may substitute ${\cal R}=-\kappa{\cal T}$ in the argument of ${\cal S}(\r)$ function in  the second term in the relation (\ref{eight}):
\begin{equation}
{\cal R}_{\mu\nu}=\kappa \left ({\cal T}_{\mu\nu}-\frac{1}{2}{\cal T}g_{\mu\nu}\right ) -\beta \left (-{\cal T}_{\mu\nu}\dot{{\cal S}}({\cal T})+\frac{1}{2}{\cal S}({\cal T})g_{\mu\nu}\right )+{\cal O}(\beta^2)
\label{nine}
\end{equation}
where $\dot{{\cal S}}=\frac{d{\cal S}({\cal T})}{d{\cal T}}$. And finally for the other terms in  the Raychaudhuri's equation we substitute $\ln f'=-\beta\dot{{\cal S}}/\kappa+{\cal O}(\beta^2)$.
Thus equation (\ref{ohoy}) reads as:
\[
\left [ \kappa \left ({\cal T}_{\mu\nu}-\frac{1}{2}{\cal T}g_{\mu\nu}\right ) -\beta \left (-{\cal T}_{\mu\nu}\dot{{\cal S}}({\cal T})+\frac{1}{2}{\cal S}({\cal T})g_{\mu\nu}\right )\right ] u^\mu u^\nu+
\]
\begin{equation}
3\beta \left ( \frac{\tilde\theta}{\kappa}\frac{d\dot{{\cal S}}}{d\tau}+\frac{1}{2\kappa}\frac{d^2\dot{{\cal S}}}{d\tau^2}-\frac{1}{6\kappa}\nabla_\alpha\nabla^\alpha{\cal S} \right ) \ge 0
\label{ten}
\end{equation}
up to this order.

Let us consider the special case $f=\r+\beta\r^n$(which has cosmological applications especially for negative $n$\cite{car}),  and consider a dust with ${\cal T}_{\mu\nu}=\rho u_\mu u_\nu$. The above equation then reads as:
\[
\rho+\beta (-\kappa)^{n-2}\left (  (2n-1)\kappa\rho^n +6n(n-1)\tilde{\theta}\rho^{n-2} \frac{d\rho}{d\tau} +
3n(n-1)(n-2)\rho^{n-3}\left ( \frac{d\rho}{d\tau}\right )^2
\right .
\]
\begin{equation}
\left . +
3n(n-1) \frac{d^2\rho}{d\tau^2} -\nabla_\alpha\nabla^\alpha\rho^n \right )\ge 0
\end{equation}

For the Einstein's theory in which $\beta=0$ this energy condition coincides with the physical condition $\rho \ge 0$. But for $\beta\ne 0$ even having the physical condition $\rho \ge 0$, the above condition is not  necessarily satisfied, so the congruence is not necessarily convergent.

Another suitable choice is to consider the cosmological constant. The corresponding energy--momentum tensor is ${\cal T}_{\mu\nu}=\frac{\Lambda}{\kappa} g_{\mu\nu}$. The convergence condition of the congruence is now:
\begin{equation}
-\Lambda+\frac{\beta\Lambda}{\kappa}\dot{\cal {S}}-\frac{\beta}{2}{\cal S} \ge 0
\end{equation}
For the special case $f=\r+\beta\r^n$ we get:
\begin{equation}
-\Lambda+\frac{\beta}{4}(-4\Lambda)^n(n-2) \ge 0
\end{equation}
It is clear from the above relation that for Einstein's theory ($\beta=0$) the convergence condition is $\Lambda\le 0$, as it should be. On the other hand there is a special case ($n=2$) for which the same condition is achieved. It is interesting to note that it is possible to choose $n$ and $\beta$ such that even for $\Lambda\ge 0$ one gets converging congruence.

We can conclude now that Einstein's general relativity and Pallatini $f(\r)$ theory can completely disagree on attractive character of gravity. For example for a dust with $\rho\ge 0$ the first gives attraction while the second can give repulsion. Also for a positive cosmological constant, the first gives repulsion, while the second can give attraction.

\textbf{Acknowledgement} This work is partly supported by a grant from University of Tehran and partly by a grant from Center of Excellence of Department of Physics on the Structure of Matter.

\end{document}